\documentclass[11pt,fleqn]{article}

\usepackage{amsmath,amssymb,amsthm,enumerate,cite}

\newcommand{\p}{\partial}
\newcommand{\const}{\mathop{\rm const}\nolimits}

{\theoremstyle{definition}
\newtheorem{example}{Example}
}

\begin{document}

\par\noindent {\LARGE\bf
Construction of potential systems for systems of PDEs with multi-dimensional spaces of conservation laws
\par}
{\vspace{4mm}\par\noindent {\bf N.M. Ivanova
} \par\vspace{1mm}\par}
{\vspace{1mm}\par\noindent {\it
Institute of Mathematics of NAS of Ukraine,
3 Tereshchenkivska Str., 01601 Kyiv, Ukraine\\
}}
{\noindent \vspace{2mm}{\it
e-mail: ivanova@imath.kiev.ua
}\par}

{\vspace{2mm}\par\noindent\hspace*{8mm}\parbox{140mm}{\small
In this paper we consider generalization of procedure of construction of potential systems for systems of partial differential equations
with multidimensional spaces of conservation laws.
More precisely, for construction of potential systems in cases when dimension of the space of
local conservation laws is greater than one, instead of using only basis conservation laws we use their arbitrary
linear combinations being inequivalent with respect to equivalence group of the class of systems or symmetry group of the fixed system.
It appears that the basis conservation laws can be equivalent with respect to groups of symmetry or equivalence transformations, or vice versa,
the number of independent in this sense linear combinations of conservation laws can be grater than the dimension of the space of conservation laws.
The first possibility leads to an unnecessary, often cumbersome, investigation of equivalent systems, the second one makes possible missing
a great number of inequivalent potential systems. Examples of all these possibilities are given.
}\par\vspace{2mm}}

\section{Introduction}

When one tries to investigate a physical process, knowledge of symmetry properties of the modelling system of PDEs
can be very useful for understanding the behavior of solutions of the model.
Thus, e.g., group invariance property of a system of differential equations allows to generate new solutions from the known ones,
to construct conservation laws, to find wide classes of exact invariant solutions.
This becomes especially important for investigation of nonlinear models, where every single partial solution plays an important role.
Even if it does not solve any real boundary value problem, it can be used, e.g., as a testing solution for different numeric
or approximate algorithms.
Moreover, for many nonlinear systems invariant solutions are the only known solutions.
This is only one reason why any new symmetry is of great importance for systems of PDEs.
In this short note we illustrate a way of construction of bigger number of potential systems that can lead to finding new potential symmetries.

The concept of potential symmetry was introduced by
Bluman {\it at al}~\cite{Bluman&Kumei1989,Bluman&Reid&Kumei1988} in the late 80-es.
See also the related notion of quasi-local symmetry~\cite{Akhatov&Gazizov&Ibragimov1987,Akhatov&Gazizov&Ibragimov1989}.
Namely, if at least one of equations of a system of PDEs can be written in conserved form, then using it one can
introduce potential variable(s). Attaching equations containing the new potential variable(s) to the system,
one obtains a new system, nonlocally related to the initial one. Moreover, there exists a one-to-one correspondence between
the solutions of the initial and potential systems. Also, any symmetry of the potential system induces symmetry of the initial system.
(Generally speaking, the inverse statement is a bit different: symmetries of initial system induce symmetries of potential systems
or equivalence transformations in the set of potential systems corresponding to the initial system~\cite{Popovych&Ivanova2004ConsLawsLanl}.)
If the symmetry transformations of the local variables depend explicitly on the potential variable(s),
the obtained symmetry projects to a nonlocal for the initial system and is called {\it potential symmetry}.

The above procedure of finding potential symmetries has been generalized in~\cite{Popovych&Ivanova2004ConsLawsLanl}
by admitting dependence of symmetries on potentials associated to several conservation laws simultaneously.
Below we will use attitude ``simplest" to emphasize that the potential system is constructed with usage of one conservation law only.
Here we use slightly generalized procedure of construction of potential systems for systems of PDEs
admitting multi-dimensional spaces of conservation laws.

Before, for construction of potential symmetries only basis conservation laws were used.
However, such way does not guarantee that the obtained potential systems (and therefore, the obtained potential symmetries)
will be inequivalent with respect to local symmetry group of the initial system or local equivalence group of the class of system.
This may lead to unnecessary complicated investigation of equivalent systems with, in fact, no new result.
Or vice versa, there exists a linear combination of basis conservation laws that leads to potential system inequivalent to ``basis'' ones.
In such case, considering only potential systems constructed with basis conservation laws one can ``loose''
some of potential symmetries.

\section{Basic notions on conservation laws and potential symmetry}

For simplicity here we consider the case of simplest potential systems for
systems of $(1+1)$-dimensional equations only. Note that all below statements and notions can be
easily generalized to $n$-dimensional case and to general potential systems constructed using several conservation laws simultaneously.

Let~$\mathcal{L}$ be a system~$L(t,x,u_{(\rho)})=0$ of $l$ PDEs $L^1=0$, \ldots, $L^l=0$
for the unknown functions $u=(u^1,\ldots,u^m)$
of the independent variables~$t$ and~$x$.
Here $u_{(\rho)}$ denotes the set of all partial derivatives of the functions $u$
of order not greater than~$\rho$, including $u$ as the derivatives of the zero order.

Roughly speaking, a {\em conservation law} of the system~$\mathcal{L}$ is a divergence expression
\begin{equation}\label{conslaw}
D_tT(t,x,u_{(r)})+D_xX(t,x,u_{(r)})=0
\end{equation}
which vanishes for all solutions of~$\mathcal{L}$.
Here $D_t$ and $D_x$ are the operators of total differentiation with respect to $t$ and $x$, respectively.
The differential functions $T$ and $X$ are correspondingly called a {\em density} and a {\em flux} of the conservation law and
the tuple $(T,X)$ is a \emph{conserved vector} of the conservation law.

The crucial notion of the theory of conservation laws is one of equivalence and triviality of conservation laws.
Two conserved vectors $(T,X)$ and $(T',X')$ are {\em equivalent} if
there exist functions~$\hat T$, $\hat X$ and~$H$ of~$t$, $x$ and derivatives of~$u$ such that
$\hat T$ and $\hat X$ vanish for all solutions of~$\mathcal{L}$~and
$T'=T+\hat T+D_xH$, $X'=X+\hat X-D_tH$.
A conserved vector is called {\em trivial} if it is equivalent to the zeroth vector.

The notion of linear dependence of conserved vectors is introduced in a similar way.
Namely, a set of conserved vectors is {\em linearly dependent}
if a linear combination of them is a trivial conserved vector.

Although in many simple cases conservation laws can be investigated in the above empiric framework,
for deeper analysis one often needs to consider more rigorous definitions,
that can be found, e.g., in~\cite{Zharinov1986,Popovych&Ivanova2004ConsLawsLanl,Ivanova&Popovych&Sophocleous2007Part3}.

Let the system~$\cal L$ be totally nondegenerate~\cite{Olver1993}.
Then application of the Hadamard lemma to the definition of conservation law and integrating by parts imply that
the left hand side of any conservation law of~$\mathcal L$ can be always presented up to the equivalence relation
as a linear combination of left hand sides of independent equations from $\mathcal L$
with coefficients~$\lambda^\mu$ being functions of $t$, $x$ and derivatives of~$u$:
\begin{equation}\label{CharFormOfConsLaw}
D_tT+D_xX=\lambda^1 L^1+\dots+\lambda^l L^l.
\end{equation}

Formula~\eqref{CharFormOfConsLaw} and the $l$-tuple $\lambda=(\lambda^1,\ldots,\lambda^l)$
are called the {\it characteristic form} and the {\it characteristic}
of the conservation law~$D_tT+D_xX=0$ correspondingly.

The characteristic~$\lambda$ is {\em trivial} if it vanishes for all solutions of $\cal L$.
Since $\cal L$ is nondegenerate, the characteristics~$\lambda$ and~$\tilde\lambda$ satisfy~\eqref{CharFormOfConsLaw}
for the same conserved vector~$(T,X)$ and, therefore, are called {\em equivalent}
iff $\lambda-\tilde\lambda$ is a trivial characteristic.

Any conservation law~\eqref{conslaw} of~$\mathcal{L}$ allows us to deduce the new dependent (potential) variable~$v$
by means of the equations
\begin{equation}\label{potsys1}
v_x=T,\qquad v_t=-X.
\end{equation}

In the case of single equation~$\mathcal{L}$, equations of form~\eqref{potsys1} combine into
the complete potential system since~$\mathcal{L}$ is a differential consequence of~\eqref{potsys1}.
As a rule, systems of such kind admit a number of nontrivial symmetries and so they are of a great interest.
If the transformation of some of nonlocal variables~$t$, $x$ or~$u$ depends explicitly on
variable~$v$, such symmetry is a nonlocal for the initial equation (system) and is called
{\em potential symmetry}.

\section{New potential systems}

In~\cite{Ivanova&Popovych&Sophocleous2007Part3} a new approach of choosing conservation laws for introducing potentials
in order to obtain all possible inequivalent potential systems, has been proposed.
More precisely, for construction of potential systems in cases when the dimension of the space of
conservation laws is greater than one, instead of using only basis conservation laws we propose to use their arbitrary
linear combinations being inequivalent with respect to equivalence group of the class of systems or symmetry group of the fixed system.
It is appeared that the basis conservation laws can be equivalent with respect to groups of symmetry or equivalence transformations, or vice versa,
the number of independent in this sense linear combinations of conservation laws can be grater then dimension of the space of conservation laws.
The first possibility leads to an unnecessary, often cumbersome, investigation of equivalent systems, the second one makes possible missing
a great number of inequivalent potential systems.
Below we illustrate all these three possibilities and show an example when such systems lead to new potential symmetries.

The most classical in this sense is an example of diffusion equations, for which indeed all possible inequivalent potential systems
can be constructed with usage of basis conservation laws only.
\begin{example}
Consider a class of nonlinear diffusion equations of form
\begin{equation}\label{eqDifEq}
u_t=(A(u)u_x)_x,\qquad  A(u)\ne\const.
\end{equation}
Equivalence group~$G^{\sim}_1$ of this class consists of transformations
\[
\tilde t=\varepsilon_1t+\varepsilon_4 ,\quad
\tilde x=\varepsilon_2x+\varepsilon_5 ,\quad
\tilde u=\varepsilon_3u+\varepsilon_6 ,\quad
\tilde A=\varepsilon_1^{-1}\varepsilon_2^2A,
\]
where $\varepsilon_1,\ldots, \varepsilon_6$ are arbitrary constants, $\varepsilon_1\varepsilon_2\varepsilon_3\ne0$.

It is well-known that for arbitrary value of~$A$ this equation
possesses two linearly independent conservation laws of form
\[\textstyle
D_t(u)+D_x(-Au_x)=0\quad \mbox{and}\quad D_t(xu)+D_x(-xAu_x+\int\! Adu)=0.
\]
Therefore, the most general form of potential system depending on one potential that can be constructed for~\eqref{eqDifEq} has the form
\[\textstyle
v_x=(c_1x+c_2)u,\quad v_t=c_1(xAu_x-\int\! Adu)+c_2Au_x.
\]
Depending on value of~$c_1$ (is it equal to~$0$ or not), using translation of~$x$ from equivalence group~$G^{\sim}$
and trivial scaling of the potential variable
this general system can be mapped to one of the following two inequivalent systems:
\begin{gather*}
v^1_x=u, \quad v^1_t=Au_x,\qquad \mbox{or}\\ \textstyle
v^2_x=xu, \quad v^2_t=xAu_x-\int\! Adu.
\end{gather*}
Together with potential system
\[\textstyle
v^1_x=u, \quad v^1_t=Au_x,\quad v^2_x=xu, \quad v^2_t=xAu_x-\int\! Adu,
\]
constructed with simultaneous usage of two potentials they exhaust all possible inequivalent potential systems
that can be constructed from local conservation laws of equations of class~\eqref{eqDifEq}.
\end{example}

In the second example we have a different situation: the number of inequivalent potential systems is less than the dimension of the space
of conservation laws.
Although from the physical point of view the example seems to be a bit artificial, it is an excellent illustration of such possibility.
\begin{example}
Consider a class of diffusion-convection equations of form
\[
e^{\mu\arctan x}(x^2+1)^{-3/2}u_t=(A(u)u_x)_x+e^{\mu\arctan x}(x^2+1)^{-1/2}u_x
\]
with two-dimensional space of conservation laws spanned by ones with the conserved vectors
\begin{gather*}\textstyle
(\,e^{\mu t}(x\cos t+\sin t)fu ,\ -e^{\mu t}(x\cos t+\sin t)(Au_x+hu)+e^{\mu t}\cos t\int\! Adu\,),\\ \textstyle
(\,e^{\mu t}(x\sin t-\cos t)fu, \ -e^{\mu t}(x\sin t-\cos t)(Au_x+hu)+e^{\mu t}\sin t\int\! Adu\,).
\end{gather*}

It is easy to see that under the action of equivalence transformations of time translation
and scaling of potential variable there exist only one locally inequivalent simplest potential system having the form
\begin{gather*}\textstyle
v_x=e^{\mu t}(x\cos t+\sin t)fu ,\quad v_t=e^{\mu t}(x\cos t+\sin t)(Au_x+hu)-e^{\mu t}\cos t\int\! A du.
\end{gather*}
\end{example}

At last, consider the most interesting example of the class of wave equations illustrating the possibility of construction of extra
potential systems, yielding new potential symmetries.
\begin{example}
Consider the class of wave equations
\begin{equation}\label{eqWaveEq}
 u_{tt}=(f(u)u_x)_x.
\end{equation}
Its equivalence group~$G^{\sim}_w$ consists of scaling and translation transformations of~$t$, $x$ and~$u$.

For this equation the following local conservation laws with characteristics of zero order are known (see, e.g.,~\cite{Bluman&Cheviakov2007}):
\begin{gather*}
D_t(u_t)-D_x(fu_x)=0,\quad D_t(tu_t-u)-D_x(tfu_x)=0,\\ \textstyle
D_t(xu_t)-D_x(xfu_x-\int\! fdu)=0, \quad D_t(x(tu_t-u))-D_x(t(xfu_x-\int\! fdu))=0.
\end{gather*}
Their characteristics are~$1$, $t$, $x$ and~$tx$ correspondingly. Therefore, the most general simplest potential system
can be constructed with usage of local conservation law having characteristic~$c_1tx+c_2x+c_3t+c_4$, where $c_i$ are arbitrary constants.

If $c_1\ne0$, then without loss of generality we can assume that $c_1=1$.
Using equivalence transformations $x\to x-c_3$, $t\to t-c_2$ we can reduce this characteristic to form $xt+c_4$.
Applying additionally scaling transformations we get $xt+\varepsilon$, where $\varepsilon=0,1$.
Considering similarly case $c_1=0$ we obtain the following inequivalent in this sense characteristics: $x+\varepsilon t$, $t$ and $1$,
where $\varepsilon=0,1$. In such way we get the following inequivalent simplest potential systems:
\begin{gather}
v^1_x=u_t , \quad v^1_t=fu_x , \label{sysPotSysWaveEqChar1}\\
v^2_x=tu_t-u , \quad v^2_t=tfu_x , \label{sysPotSysWaveEqChart}\\ \textstyle
v^3_x=xu_t+\varepsilon(tu_t-u) , \quad v^3_t=xfu_x-\int\! fdu+\varepsilon tfu_x , \label{sysPotSysWaveEqCharx+ct}\\ \textstyle
v^4_x=x(tu_t-u)+\varepsilon u_t , \quad v^4_t=t(xfu_x-\int\! fdu)+\varepsilon fu_x . \label{sysPotSysWaveEqCharxt+c}
\end{gather}

Classification of symmetries of potential systems~\eqref{sysPotSysWaveEqChar1}, \eqref{sysPotSysWaveEqChart},
\eqref{sysPotSysWaveEqCharx+ct}$|_{\varepsilon=0}$ and \eqref{sysPotSysWaveEqCharxt+c}$|_{\varepsilon=0}$
is considered in~\cite{Bluman&Cheviakov2007},
while potential systems
\begin{gather}\textstyle
v_x=xu_t+tu_t-u , \quad v_t=xfu_x-\int\! fdu+tfu_x , \label{sysPotSysWaveEqCharx+t}\\ \textstyle
v_x=x(tu_t-u)+u_t , \quad v_t=t(xfu_x-\int\! fdu)+ fu_x , \label{sysPotSysWaveEqCharxt+1}
\end{gather}
are new and can lead to new potential symmetries.
In particular,
system~\eqref{sysPotSysWaveEqCharxt+1} gives potential symmetries for equations~\eqref{eqWaveEq}
if and only if $f=1\!\!\mod G^{\sim}_w$. The corresponding potential algebra has the form
\begin{gather*}
\Big\langle
\frac1{x^2-t^2}(t\p_t-x\p_x),\,
\left(t+\frac{2x}{x^2-t^2} \right)\p_t+\left(x-\frac{2t}{x^2-t^2} \right)\p_x,\\
-\frac14(3tx^2+4x+t^3)\p_t-\frac14(x^3+3t^3x+4t)\p_x+\left(v+\frac12(t^2+x^2)u\right)\p_u+\\
\left((1+2tx+t^2x^2)u-\frac12(t^2+x^2)v\right)\p_v,\,
\mu\p_u+\phi\p_v
\Big\rangle,
\end{gather*}
where $\mu=\mu(t,x)$ and $\phi=\phi(t,x)$ satisfy the following system of linear equations
\[
\phi_t=(xt+1)\mu_x-t\mu,\quad \phi_x=(xt+1)\mu_t-x\mu.
\]
Similarly one can prove that equation~\eqref{eqWaveEq} with~$f=1$ admits potential symmetries
associated with potential system~\eqref{sysPotSysWaveEqCharx+t}.
\end{example}

\subsection*{Acknowledgements}
The author is grateful to the organizers of the Conference ``Similarity: generalizations, applications and open problems''
for invitation and financial support.
This research was supported by the Grant of the President of Ukraine for young scientists
(project number GP/F26/0005).

\end{document}